\documentclass[referee]{aa}
\usepackage{amsmath,amssymb,graphicx,url}
\newcommand{\bd}[2]{{#1}_{\scriptscriptstyle {\rm #2}}}
\newcommand{\rrad}{\bd{\rho}{R}}
\newcommand{\rmat}{\bd{\rho}{M}}
\newcommand{\rlam}{\bd{\rho}{{\Lambda}}}
\newcommand{\Orado}{\bd{\Omega}{R,0}}
\newcommand{\Omato}{\bd{\Omega}{M,0}}
\newcommand{\Olamo}{\bd{\Omega}{{\Lambda},0}}
\newcommand{\too}{\bd{t}{0}}
\newcommand{\Ho}{\bd{H}{0}}
\newcommand{\tauo}{\bd{\tau}{0}}
\newcommand{\hoo}{\bd{h}{0}}
\newcommand{\rcrito}{\bd{\rho}{crit,0}}
\newcommand{\dL}{\bd{d}{L}}
\newcommand{\mfid}{\bd{m}{fid}}
\newcommand{\Dfid}{\bd{{\cal D}}{fid}}
\newcommand{\smallfrac}[2]{\textstyle{\frac {#1}{#2}}}
\newcommand{\muo}{\bd{\mu}{0}}
\newcommand{\curlyLcrit}{\bd{\cal L}{crit}}
\newcommand{\Lcrit}{\bd{L}{crit}}
\newcommand{\Lmax}{\bd{L}{max}}
\newcommand{\Lmin}{\bd{L}{min}}
\newcommand{\lpeak}{\bd{\ell}{peak}}
\begin{document}

\title{Decaying Dark Energy in Higher-Dimensional Gravity}

\author{J.M. Overduin\inst{1} \and P.S. Wesson\inst{2} \and B. Mashhoon\inst{3}}

\institute{Gravity Probe B, Hansen Experimental Physics Laboratory,
   Stanford University,\\Stanford, CA 94305-4085
\and Astronomy Group, Department of Physics, University of Waterloo,\\
   Waterloo, Ontario N2L 3G1
\and Department of Physics and Astronomy, University of Missouri-Columbia,\\
   Columbia, MO 65211}


\abstract{}{We use data from observational cosmology to put constraints on
higher-dimensional extensions of general relativity in which the effective
four-dimensional dark-energy density (or cosmological ``constant'') decays
with time.}{In particular we study the implications of this decaying dark
energy for the age of the universe, large-scale structure formation, big-bang
nucleosynthesis and the magnitude-redshift relation for Type~Ia supernovae.}
{Two of these tests (age and the magnitude-redshift relation) place modest
lower limits on the free parameter of the theory, a cosmological length scale
$L$ akin to the de~Sitter radius.  These limits will improve if experimental
uncertainties on supernova magnitudes can be reduced around $z\sim1$.}{}


\maketitle

\section{Introduction}

In standard general relativity, dark energy is interchangeable with 
Einstein's cosmological constant $\Lambda$ and the dark-energy density
$\rlam=\Lambda/(8\pi G)$ is constant.  Observation tells us that,
at the present time, this latter quantity is nonzero but many orders of
magnitude smaller than expected based on calculations in quantum field
theory.  This mismatch has led some theorists to look at alternatives
to standard general relativity in which $\Lambda$ is a dynamical parameter
whose value is not constant but might have decayed from large values in
the early universe to those we see today (Overduin \& Cooperstock 1998).

Here we look at one such alternative, the minimal extension of Einstein's
theory to manifolds with one additional noncompact spacelike dimension
(Overduin \& Wesson 1997).  Higher-dimensional gravity has been shown to
be compatible with solar-system tests including the classical tests of
general relativity (Kalligas et al. 1995, Liu \& Overduin 2000) and
experimental limits on violations of the equivalence principle
(Overduin 2000).  We wish to check here whether it is also consistent
with basic cosmological tests that are sensitive to the $\Lambda$ term.
An exhaustive treatment of all cosmological data is beyond the scope of
this introductory study. 

Mashhoon \& Wesson (2004) have shown that a gauge transformation in
higher-dimensional gravity converts $\Lambda$ from a constant of nature
to a gauge-dependent measure of four-dimensional vacuum-energy density.
Under reasonable assumptions (conformal flatness and geodesic motion in
five dimensions), $\Lambda$ decays {\em exponentially\/} with cosmic time.
We investigate the cosmological implications of this kind of vacuum decay
by solving numerically for the age $t(z)$ of the universe as a function of
redshift.  The bulk of the vacuum decays takes place near $z\sim1$, too
late to affect big-bang nucleosynthesis (BBN) or large-scale structure
formation (LSS).  Lower limits on the age of the universe, however, put
weak constraints on the primary free parameter of the theory
(a de~Sitter-like length parameter $L$), and these bounds are improved
somewhat by data on the magnitude-redshift relation for Type~Ia supernovae.
Taken together with the solar-system tests, we tentatively conclude that
available astrophysical data are consistent with a universe with one
(or more) extra dimensions.

\section{Higher-Dimensional Cosmology}

A starting point for cosmological investigations in 5D theory is
the metric in canonical form (Mashhoon et al. 1994):
\begin{equation}
dS^2 = \frac{\ell^2}{L^2} \left[ g_{\mu\nu} (x^{\lambda},\ell)
   dx^{\mu} dx^{\nu} \right] - d\ell^2 \; , 
\label{5dmetric}
\end{equation}
where $L$ is a constant with dimensions of length (akin to the de~Sitter
radius $L_d=\sqrt{3/\Lambda}$ in standard cosmology).  The 5D line element
contains the 4D one:
\begin{equation}
ds^2 = g_{\mu\nu} (x^{\lambda},\ell) dx^{\mu}dx^{\nu} \; .
\label{4dmetric}
\end{equation}
There is no loss of generality to this point; five available degrees of
coordinate freedom have been used to set the electromagnetic potentials
($g_{4\mu}$) to zero and to set the scalar potential ($g_{44}$) to a constant
in Eq.~(\ref{5dmetric}).  It is, however, necessary to retain
$\ell$-dependence in the 4D metric tensor in order to preserve this
generality (Overduin \& Wesson 1997).

Under the restriction to 5D conformal flatness, and the natural assumption
that all test particles (massive as well as massless) move along null geodesics 
in 5D (i.e., $dS^2=0$), Mashhoon \& Wesson (2004) have shown that 
$\Lambda$ in 4D drops exponentially with proper time $s$:
\begin{equation}
\Lambda = \frac{3}{L^2} \frac{1}{(1 - e^{\mp s/L})^2} \; .
\label{Lambda-s}
\end{equation}
Physically, this variation arises because we require the 5D field equations
to satisfy general covariance in five, not four dimensions.  The canonical
metric~(\ref{5dmetric}) is invariant with respect to translations along the
$\ell$-axis, so $\ell$- or gauge-dependence then necessarily appears
in the 4D field equations.  We have used the 4D metric to re-express this
dependence in terms of proper time $s$ rather than $\ell$.  There are two
cases: in the first (`$-$' sign in the exponent), $\Lambda$ decays
asymptotically to the small finite value $3/L^2$ as $s\rightarrow\infty$,
while in the second (`$+$' sign) it vanishes in this limit.
Measurements tell us that $\Lambda$ is small at present, but are not precise
enough to discriminate between a constant value and one that is still
decaying on cosmological timescales.  Therefore we retain both possibilities
in what follows.

One way to constrain proposals of this kind is to ask what $\Lambda$ decays
{\em into}.  If matter or radiation, then strong constraints can be placed
on the theory using experimental limits on cosmic background radiation
(Overduin et al. 1993, Overduin \& Wesson 2003, 2004).  In this paper,
we will investigate the more conservative scenario in which $\Lambda$ does
{\em not\/} decay into matter or radiation, so that the evolution of the
matter density ($\rmat$) and radiation energy density ($\rrad$) proceed
just as in standard 4D cosmology.  It should be noted that such an
assumption requires, in principle, the existence of some other field to
which $\Lambda$ is coupled and into which its energy can be transferred,
in accordance with the 4D conservation law (Overduin \& Cooperstock 1998):
\begin{equation}
\nabla^{\nu} ( 8\pi G {\cal T}_{\mu\nu} - \Lambda g_{\mu\nu} ) = 0 \; .
\end{equation}

In order to compare Eq.~(\ref{Lambda-s}) to observation, we need to convert
from proper time $s$ to ordinary cosmic time $t$.  A physically-motivated
argument due to Mashhoon is the following: translations along the $\ell$-axis
do not only introduce gauge-dependence into $\Lambda$; they also give rise to
an apparent ``fifth force'' in the equations of motion (Mashhoon \& Wesson
2004, Wesson 2005).  This force can however be made to vanish by an
appropriate choice of affine parameter, so making the motion geodesic in the
usual 4D sense (Seahra \& Wesson 2001).  This choice can be shown to lead
to the following relation between $s$ and $t$, assuming that the motion in
5D is null, that the galaxies are comoving in 4D spacetime, and that
$dt/ds>0$ over $0<s<\infty$:
\begin{equation}
e^{\pm s/L} = 1 \pm (1/\alpha) e^{t/L} \; .
\label{s-t}
\end{equation}
Here $\alpha$ (expressed this way for later convenience) is an unknown
positive constant whose value is to be fixed by cosmological boundary
conditions, and the signs are to be read in the same way as in
Eq.~(\ref{Lambda-s}); i.e., the upper (`$+$') sign here corresponds to the
upper (`$-$') sign there.  Putting Eq.~(\ref{s-t}) into the decay
law~(\ref{Lambda-s}), we find that:
\begin{equation}
\Lambda(t) = \frac{3}{L^2} \left( 1 \pm \alpha e^{-t/L}
   \right)^2 \; ,
\label{Lambda-t}
\end{equation}
where the sign order again corresponds to that in Eqs.~(\ref{Lambda-s})
and (\ref{s-t}).

Eq.~(\ref{Lambda-t}) provides the starting-point for our investigation.
Motivated by the 5D results, we study dark-energy decay of the functional
form given by Eq.~(\ref{Lambda-t}) in the context of 4D cosmology from
$t=0$ at the big bang to $t\rightarrow\infty$.  This kind of exponential
dark-energy decay appears to be unique in the literature (Overduin \&
Cooperstock 1998).  In the limit $t\rightarrow0$, we observe that $\Lambda$
originates with a finite value of $3(1\pm\alpha)^2/L^2$ and decays
asymptotically toward $3/L^2$ as $t\rightarrow\infty$.  Motivated by the
cosmological-constant problem, we might expect very large values of $\alpha$
with steep drop-offs at early times.  We shall find, however, that $\alpha$
is typically within a few orders of magnitude of unity, rendering departures
from standard 4D cosmology rather mild.

\section{Decaying Dark Energy}

Evaluating Eq.~(\ref{Lambda-t}) at the present time $\too$, restoring physical
units and expressing the results in terms of the critical density, we obtain:
\begin{equation}
\Olamo = \frac{\Lambda(\too) c^2}{3 \Ho^2} = \left( \frac{c}{\Ho L} 
   \right)^{\!\!2} \!\! \left( 1 \pm \alpha e^{-c\,\too/L} \right)^2 \; ,
\label{Olam-to}
\end{equation}
where $\Ho$ is the present value of Hubble's parameter.  It will be
convenient to rescale all dynamical parameters in dimensionless terms,
so we define:
\begin{equation}
\tau \equiv \Ho\ t \;\;\; , \;\;\;
{\cal L} \equiv \Ho L/c \;\;\; , \;\;\;
{\cal H} \equiv H/\Ho \; .
\label{defs}
\end{equation}
In terms of these quantities Eq.~(\ref{Olam-to}) determines the age of the
universe $\tauo\equiv\Ho\,\too$ in terms of the two free parameters $\alpha$
and ${\cal L}$ as follows:
\begin{equation}
\tauo = {\cal L} \ln \left( \frac{\pm\alpha}{{\cal L}\sqrt{\Olamo} - 1}
   \right) \; .
\label{eq-tauo}
\end{equation}
We note from Eq.~(\ref{eq-tauo}) that the two cases corresponding to the
`$+$' and `$-$' signs are separated by a ``critical'' case with
$\curlyLcrit\equiv1/\sqrt{\Olamo}$.  For values of ${\cal L}>\curlyLcrit$
we must use the `$+$' solution, while the `$-$' solution is operative if
${\cal L}<\curlyLcrit$.  [An alternative solution, corresponding to the
negative root of Eq.~(\ref{Olam-to}), is also available in
principle, since Eq.~(\ref{Lambda-t}) has three possible square roots.
However, the root that results in the alternative solution should be excluded
as it does not have the same limit as Eq.~(\ref{Lambda-t}) for
$t\rightarrow\infty$.  This explains why the alternative solution is not taken
into consideration here.] In the limit ${\cal L}\rightarrow\curlyLcrit$
precisely, it is apparent from Eqs.~(\ref{Olam-to}) with the
definitions~(\ref{defs}) that $\alpha=0$ and the theory goes over to
standard cosmology with $\Lambda=$const.  The theory also goes over to
standard cosmology in the limit $L\rightarrow\infty$, which from
Eq.~(\ref{Olam-to}) means that $\alpha$ must go as $L^2$ for large $L$,
as we will confirm numerically below.  The physical length corresponding
to $\curlyLcrit$ is just the de~Sitter radius of standard cosmology,
$\Lcrit=c/(\Ho\sqrt{\Olamo})$, which takes the value $\Lcrit=4.9$~Gpc for
WMAP values of $\Ho$ and $\Olamo$ (Spergel et al. 2003).

We now fix the value of $\alpha$ by requiring consistency between
Eq.~(\ref{eq-tauo}) and the age of the universe $\tauo$ as obtained
by numerical integration of the Friedmann-Lema\^{\i}tre equation:
\begin{equation}
\tau(z) = \tauo - \int_0^z \frac{dz^{\prime}}{(1+z^{\prime}) \,
    {\cal H}[z^{\prime},\tau(z^{\prime})]} \; .
\label{t-de}
\end{equation}
Here $z$ is redshift and the standard expression for Hubble's parameter
is modified following Eqs.~(\ref{Lambda-t}) and (\ref{Olam-to}) so that:
\begin{eqnarray}
{\cal H}[z,\tau(z)] & = & \bigg[ \Omato (1+z)^3 + (1-\Omato-\Olamo)(1+z)^2
   \nonumber \\
   & & \left. + {\cal L}^{-2} \! \left(1 \pm \alpha 
   e^{-\tau(z)/{\cal L}}\right)^{\!2 } \right]^{1/2} \; .
\label{newHub}
\end{eqnarray}
Consistency is to be enforced by putting the boundary condition
$\tau(\infty)=0$ into Eq.~(\ref{t-de}); i.e., for each value of $L$ we
solve numerically for a value of $\alpha$ satisfying:
\begin{equation}
\int_0^{\infty} \frac{dz^{\prime}}{(1+z^{\prime}) \,
   {\cal H}[z^{\prime},\tau(z^{\prime})]} = {\cal L}
   \ln \left( \frac{\pm\alpha}{{\cal L}\sqrt{\Olamo} - 1} \right) \; .
\end{equation}
We adopt WMAP values for the present values of the density parameters,
assuming spatial flatness: $\Omato=0.135/\hoo^2=0.27$
(with $\hoo\equiv\Ho/100$~km~s$^{-1}$~Mpc$^{-1}=0.71$) and
$\Olamo=1-\Omato=0.73$ (Spergel et al. 2003).  That leaves us with only one
adjustable parameter in the theory: the de~Sitter-like length $L$.

Fig.~1 (top) shows how this cosmological consistency requirement
produces values of $\alpha$ between approximately 0.1 and 1000 for a range of
$L$-values with $L>5$~Gpc.
\begin{figure}
\resizebox{\hsize}{!}{\includegraphics{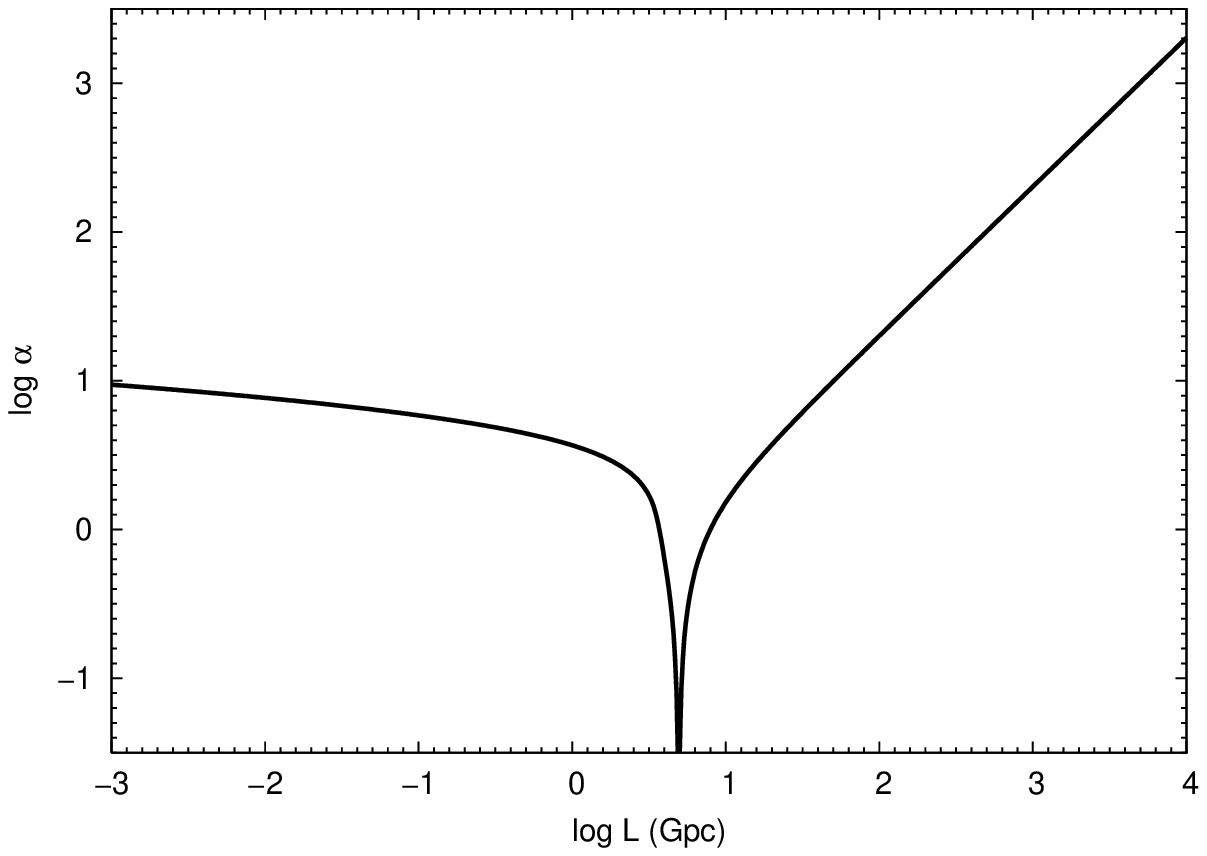}}
\resizebox{\hsize}{!}{\includegraphics{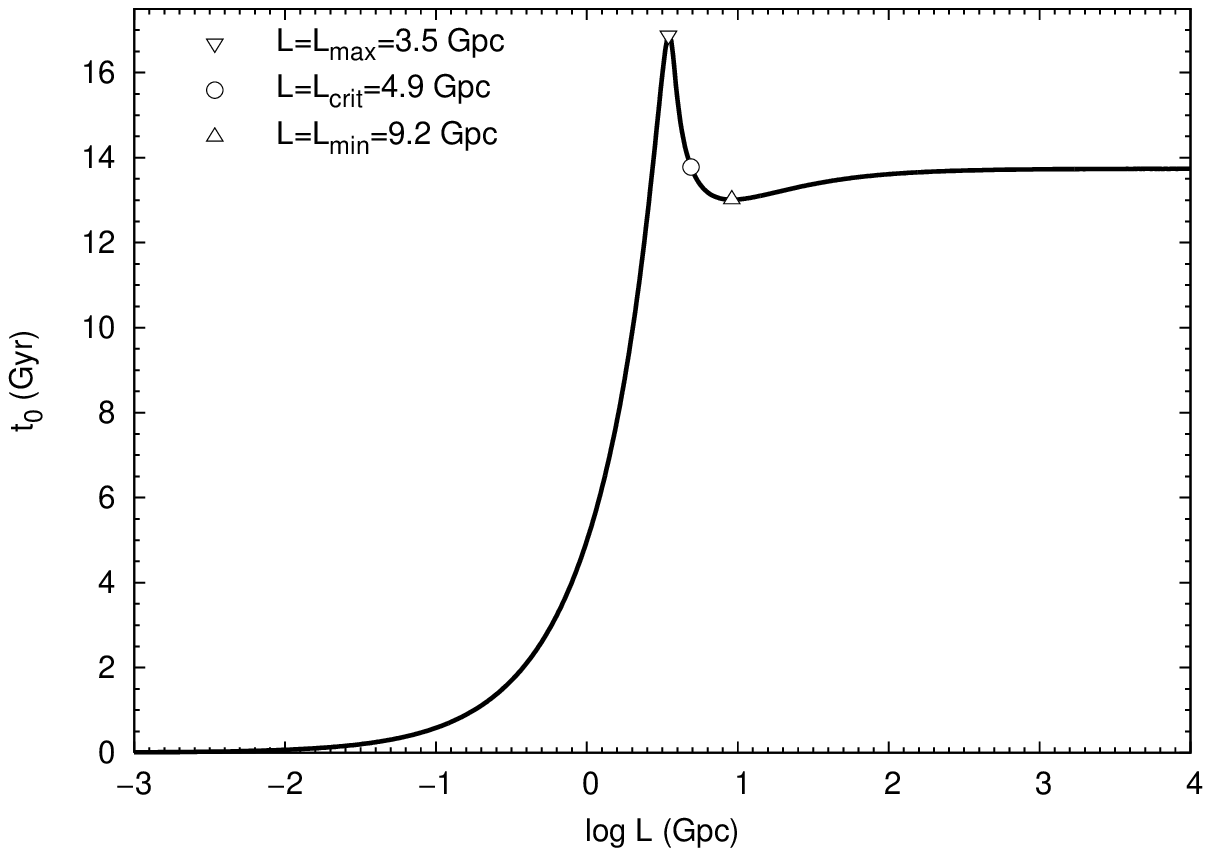}}
\caption{(Top) Values of the constant $\alpha$ as a function of
   $L$, assuming WMAP values of $\Omato$ and $\Olamo$. (Bottom)
   Age of the universe as a function of $L$ for the same values of $\Omato$
   and $\Olamo$.  Labelled points correspond to the critical case
   separating the two classes of solutions of Eq.~(\ref{eq-tauo}), and to
   universes of maximum and minimum age in the two regimes (see text for
   discussion)}
\label{alphaAgeFigs}
\end{figure}
As expected, $\alpha\rightarrow0$ in the limit $L\rightarrow\Lcrit$
and $\alpha\propto L^2$ for large $L$.

{Fig.~1 (bottom) shows the corresponding age of the universe
$\too=\tauo/\Ho$ [as obtained from either Eq.~(\ref{eq-tauo}) or
Eq.~(\ref{t-de}) in the limit $z\rightarrow\infty$] for the same range of
values of $L$ as in Fig.~1 (top).  The shape of this plot can be
understood physically as follows: The age of the universe goes over to that of
a standard flat $\Lambda$CDM model with $\Omato=0.27$ (i.e., 13.7~Gyr) in both
the ``critical'' case $L=\Lcrit$ and the limit $L\rightarrow\infty$, as
expected.  To the right and left of the critical case are the `$+$' and `$-$'
regimes in Eq.~(\ref{Lambda-t}) respectively.  The presence of a local minimum
and a maximum here is due to the $L$-dependence of the dark-energy density:
the behavior near these extrema is dominated by the exponential term
$\pm\alpha\exp(-ct/L)$ in Eq.~(\ref{Lambda-t}), whose magnitude in this
region is comparable to unity.

To understand why the age
of the universe in this theory is always lower than that of standard
cosmology for $L>\Lcrit$, whereas it can climb somewhat above this value
for $L<\Lcrit$, we recall that the dark-energy density is always pinned at
its observed value at present.  In the `$+$' regime ($L>\Lcrit$), it can
only grow in the past direction relative to its value in standard cosmology,
increasing the total density of matter plus dark energy.  Since the square
of the expansion rate is proportional to total density (from the Friedmann
equation) the expansion rate goes up and the expansion timescale goes down.
Hence, for given values of $\Omato$ and $\Olamo$ the age of the universe in
this theory is always smaller than that in standard cosmology for $L>\Lcrit$.
In the `$-$' regime, by contrast, there is a range of $L$-values for which
the dark-energy density in the past direction is slightly {\em lower\/} than
that in the equivalent standard cosmology (i.e., with the same boundary
conditions); hence one obtains a slightly older universe in this region.

These departures from standard cosmology can be significant: for $L>\Lcrit$
the age of the universe drops from 13.7 to a local minimum of 13.0~Gyr at
$\Lmin=9.2$~Gpc, while for $L<\Lcrit$ it climbs to a possible maximum of
16.9~Gyr at $\Lmax\equiv3.5$~Gpc, before dropping rapidly toward zero as
$L\rightarrow0$ (assuming WMAP values of $\Omato$ and $\Olamo$ as usual).
This curve is sufficient to set a robust lower bound $L>2.2$~Gpc from the
fact that the age of the oldest observed stars in globular clusters are at
least 11~Gyr (Wanajo et al. 2002, Schatz et al. 2002).  The range of
acceptable $L$ values could be further constrained by other limits on
the age of the universe from a variety of observations, including upper
bounds (though the latter are necessarily less robust since we may not
be able to see the oldest members of any target population).  In any 
case we will obtain stronger constraints below.

As a first step we use Eq.~(\ref{t-de}) to compute the age $t(z)=\tau(z)/\Ho$
as a function of redshift.  The presence of explicit time-dependence under
the integral sign makes this a difficult equation analytically, but it can
be solved numerically.  Results are shown in Fig.~2 (top).
\begin{figure}
\resizebox{\hsize}{!}{\includegraphics{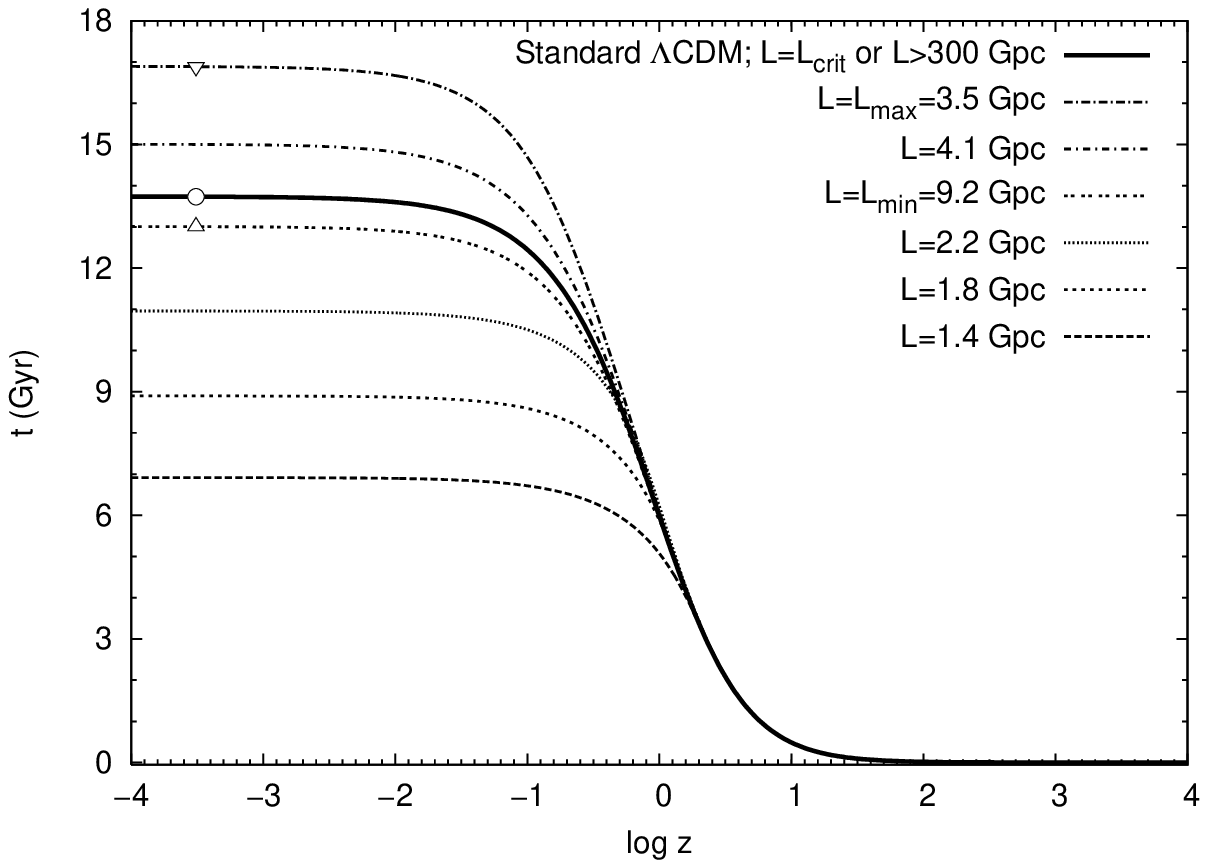}}
\resizebox{\hsize}{!}{\includegraphics{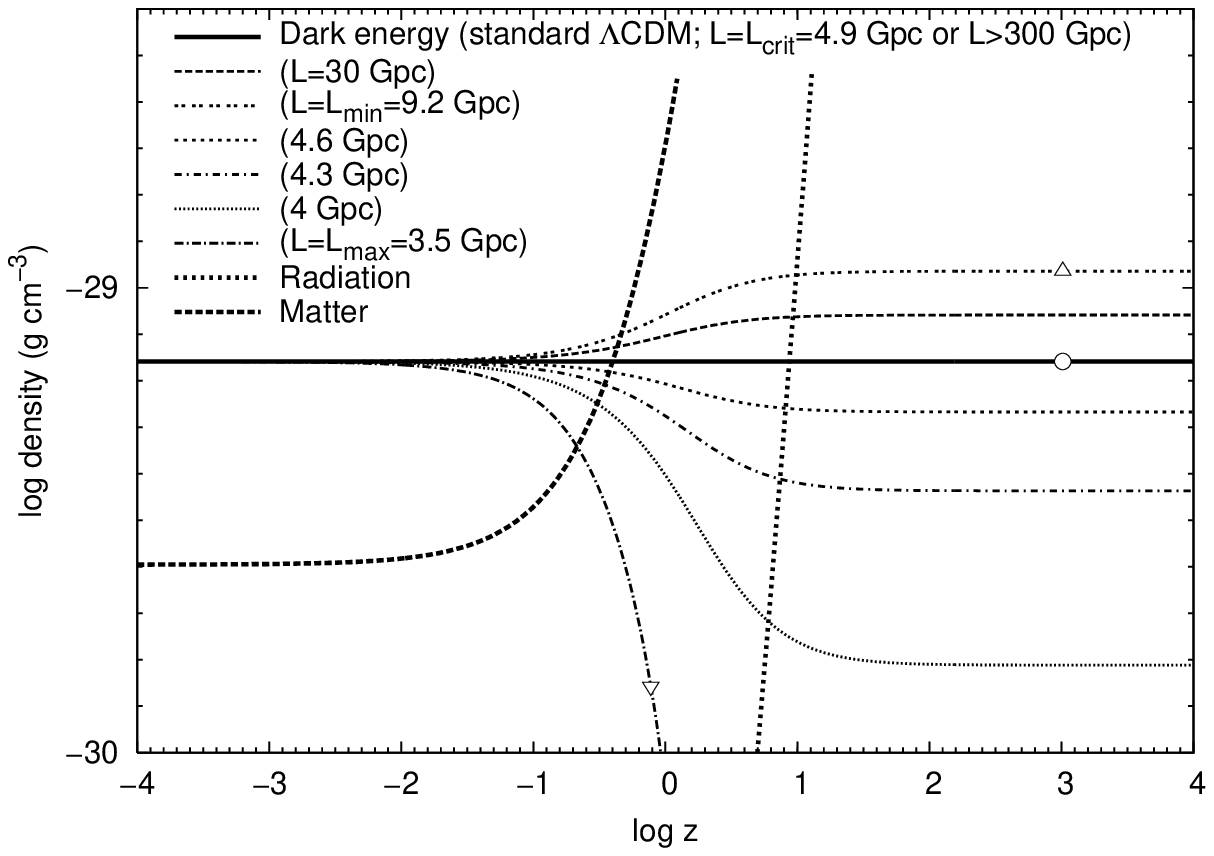}}
\caption{(Top) Age of the universe as a function of redshift for
   various values of $L$, assuming WMAP values of $\Omato$ and $\Olamo$.
   (Bottom) Densities of radiation, matter and dark energy as a
   function of redshift for various values of $L$, assuming the COBE value of
   $\Orado$ and WMAP values of $\Omato$ and $\Olamo$.  The three
   labelled points correspond to universes with critical, maximum and minimum
   values of $L$, as in Fig.~1 (bottom).}
\label{evolDensityFigs}
\end{figure}
The $L=\Lcrit, L=\Lmin$ and $L=\Lmax$ cases are again indicated together
with a representative sampling of other models.

The quantity $t(z)$ plotted in Fig.~2 (top) is helpful in
checking whether the theory is consistent with the formation of LSS by
gravitational instability in the early matter-dominated era and BBN at the end
of the radiation-dominated era.  For these processes to work successfully, the
density of dark energy must not rise so quickly in the past direction that
it becomes comparable to or greater than the matter density during LSS
formation, or comparable to or greater than the radiation energy density
during BBN.  These latter two quantities are given by:
\begin{eqnarray}
\rrad(z) & = & \rcrito \, \Orado (1+z)^4 \nonumber \\
\rmat(z) & = & \rcrito \, \Omato (1+z)^3 \; ,
\end{eqnarray}
where $\rcrito=3\Ho^2/(8\pi G)=1.88\times10^{-29}\hoo^2$~g~cm$^{-3}$ is the
critical density and $\Orado=4.17\times10^{-5}/\hoo^2$ from COBE data plus
standard neutrino physics (Overduin \& Wesson 2003).  By comparison, the
density of dark energy is given by Eq.~(\ref{Lambda-t}) as:
\begin{equation}
\rlam(z) = \rcrito \, {\cal L}^{-2} \left(1 \pm \alpha
   e^{-\tau(z)/{\cal L}} \right)^2 \; ,
\label{densities}
\end{equation}
with $\tau(z)$ as shown in Fig.~2 (top).

Inserting $\tau(z)$ into the dark-energy density~(\ref{densities}), and 
comparing with the matter and radiation energy densities, we obtain the plots
shown in Fig.~2 (bottom).  A representative sampling of models
with $L>3.5$~Gpc is included, with the $L=\Lcrit, \Lmin$ and $\Lmax$ cases
labelled as before.  It is clear from these plots that departures from standard
cosmology are too small to have a significant effect on either LSS or BBN.
For $L<\Lcrit$ the density of dark energy {\em drops\/} relative to that in
standard cosmology.  For $L>\Lcrit$, even the largest possible increase in
dark-energy density (corresponding to the shortest-lived universe with
$L=\Lmin$) is far too modest relative to the matter density at $z\gtrsim1$ to
interfere with structure formation, and completely negligible relative to
radiation-energy density at $z\gtrsim1000$.  Thus these tests do not place
meaningful constraints on the theory.

Fig.~2 (bottom) shows that the largest departures from standard
theory are found near $z\sim1$, raising the possibility that stronger
constraints might be obtained by use of the SNIa magnitude-redshift relation.
Supernovae are now being routinely monitored at $z\sim1$ (Riess et al. 2006),
providing a sensitive testbed for alternative theories of gravity
with time-varying dark-energy density (see for example Fukui 2006).

The magnitude-redshift formulae are derived in the Appendix.  We focus on
the magnitude residual $\Delta m(z)$, or difference in apparent magnitude
relative to a fiducial model, which we take here as the standard flat
$\Lambda$CDM model with WMAP values of $\Omato$ and $\Olamo$.  Predictions
are plotted as curves (for various values of $L$) in Fig.~3,
where they are compared with measurements for 92 medium-redshift SNIa at
$z>0.1$ by Tonry~et~al. (2003) and 23 high-redshift SNIa at $z\sim1$ as
compiled by Riess~et~al. (2006).
\begin{figure}
\resizebox{\hsize}{!}{\includegraphics{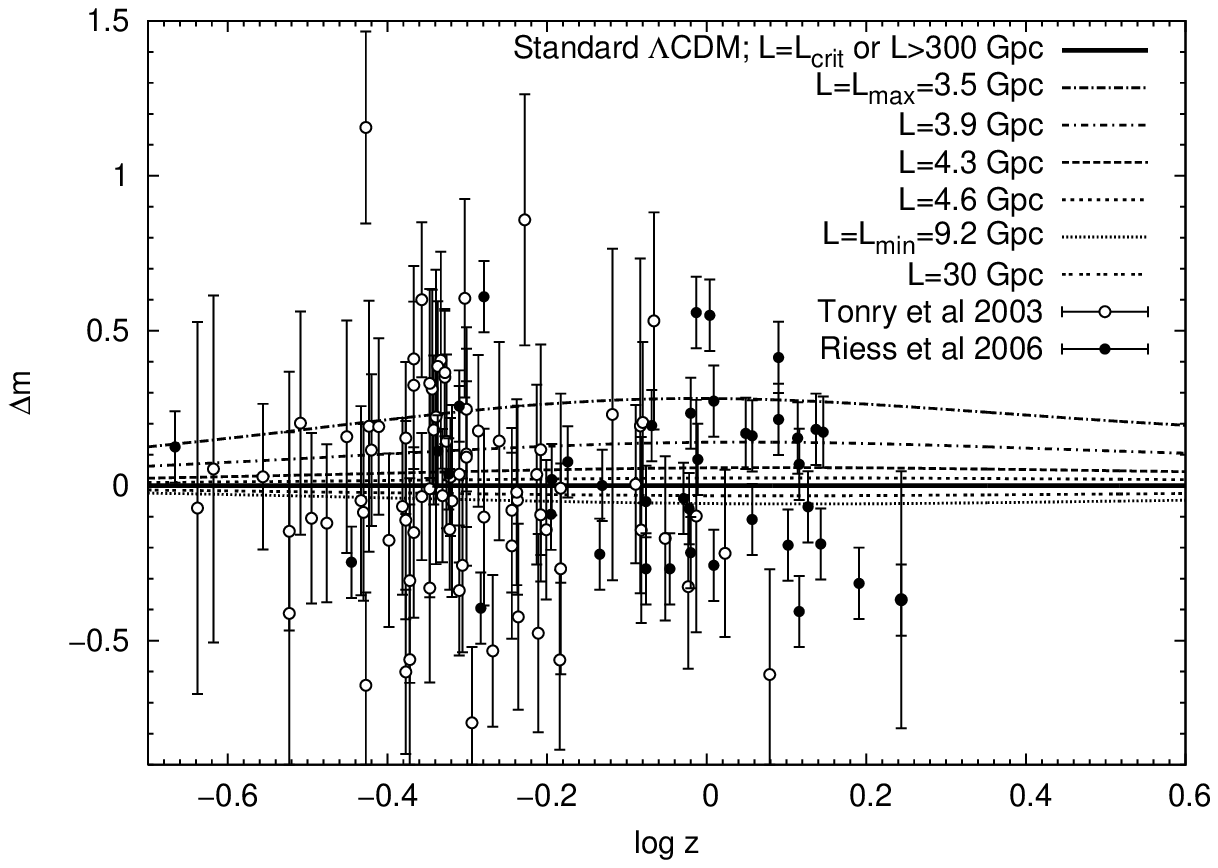}}
\caption{Magnitude-redshift relation for various values of $L$, as compared
   to observational data on Type~Ia supernovae}
\label{snFig}
\end{figure}

The heavy solid line in Fig.~3 indicates the fiducial or standard
$\Lambda$CDM model (straight line $\Delta m=0$), which overlaps with the
present theory in the case $L=\Lcrit$ and all cases with $L>300$~Gpc (to
within the precision of the plot).  For smaller values of $L$, the theory
begins to depart from standard cosmology with maximum deviations near $z\sim1$,
confirming the usefulness of SNIa as probes of the theory.  For $L$ near
(or greater than) the critical value $\Lcrit=4.9$~Gpc, the experimental
uncertainties are too large to discriminate usefully between theoretical
values of $L$.  The data are, however, good enough to disfavor smaller values
of $L$, improving significantly on the age constraint and tightening
observational bounds on the theory to $L>4.3$~Gpc.

\section{Discussion and Conclusions}

We have taken the basic extension of general relativity from 4 to 5 dimensions
and asked what observational consequences follow from its decaying
cosmological ``constant,'' or density of dark energy.  The four main
consequences involve the age of the universe, structure formation,
nucleosynthesis and the magnitude-redshift relation for Type~Ia supernovae.
There are, of course, large literatures on all four of these subjects.
We have therefore presented our results as possible departures from the
current standard model.  The theory is consistent with all four classes
of data at present.  The best way to separate 4D and 5D cosmology by
observational means in the future would appear to be by use of better
supernovae data at $z\gtrsim1$.

Another possible test of the theory might come from analysis of
the angular power spectrum of fluctuations in the cosmic microwave background
(CMB).  Qualitative considerations, however, suggest that the sensitivity of
such a test would not be competitive with those discussed above for the
kind of theory considered here.  The main feature of the CMB power spectrum
is the angular position of the first acoustic peak, $\lpeak$.  This quantity
depends only weakly on dark-energy density
[see Fig.~1 of White (1998) or the analytic approximation in Cornish (2000)].
What $\lpeak$ really measures is the {\em sum\/} of matter and dark-energy
densities, i.e. spatial curvature.  We have assumed throughout that
$\Omato+\Olamo=1$ (i.e. $k=0$), as in the standard $\Lambda$CDM model, so we
would not expect a significant shift in $\lpeak$.  Fig.~2 (bottom) shows 
that dark-energy density changes by at most $\sim\pm40$\% between $z\approx1$
and $z\approx10$ relative to standard $\Lambda$CDM cosmology.
Following Cornish (2000) the change in $\Lambda$ might shift $\lpeak$ by at
most $\sim40$\%$/35\sim1$\%, comparable to the current level of experimental
uncertainty in this parameter (Page~et~al. 2003).  Physically, the reason
why the CMB constrains this theory less strongly than SNIa is because it
probes higher redshifts where the density of matter is so much higher than 
that of dark energy that a modest change in the latter has little effect.
More detailed study is warranted, however, particularly in light of the
increase in experimental precision expected from the Planck mission.
We hope to return to this issue in future work.

These cosmological results are complementary to earlier ones based on the
classical solar-system tests and ones involving the equivalence principle.
It has been known for some time that the basic 5D extension of 4D Einstein
theory is consistent with these tests (Kalligas et al. 1995, Liu \& Overduin
2000, Overduin 2000; Wesson 2006 for review).  It is important to recall here
that standard cosmological models which are {\em curved\/} in 4D may be
smoothly embedded in models which are {\em flat\/} in 5D.
[For a review see the books by Wesson (1999,2006); for an account of the 
embeddings from an astrophysical viewpoint see Lachieze-Rey (2000).]
The data may therefore be suggesting not only that the universe has a
fifth dimension, but that its structure may be much simpler than
previously thought.

\section*{Appendix}

The apparent magnitude $m$ of a source at redshift $z$ is defined by:
\begin{equation}
m(z) = M + K(z) + 5\log\left[ \dL(z)/\mbox{10 pc} \right] \; ,
\label{appmag}
\end{equation}
where $M$ is absolute magnitude, $K(z)$ is the K-correction due to frequency
shift and the luminosity distance $\dL(z)$ is given by:
\begin{equation}
\dL(z) = \frac{c(1+z)}{\Ho\,\sigma_k} \, {\cal S}_k \! \left[ \sigma_k
   \int_0^z \frac{dz^{\prime}}{{\cal H}(z^{\prime})} \right] \; .
\end{equation}
Here the constant $\sigma_k$ and function ${\cal S}_k$ are defined so that
$\sigma_k\equiv\{\sqrt{\Omato-\Olamo-1},1,\sqrt{1-\Omato-\Olamo}\}$ and
${\cal S}_k[X]\equiv\{\sin X,X,\sinh X\}$ respectively for $k=\{+1,0,-1\}$.
Eq.~(\ref{appmag}) contains terms such as $M$, $K(z)$ and $\Ho$ that are
independent of the background cosmology and hence not of interest to us here.
We remove those terms by focusing, not on the apparent magnitude itself, but
on the difference or ``residual'' magnitude relative to a fiducial model,
which we take here to be the standard flat $\Lambda$CDM model with WMAP
values of $\Omato$ and $\Olamo$.  That is, we focus on the observational
quantity
\begin{equation}
\Delta m(z) = m(z) - \mfid(z) = 5\log\left[
   \frac{{\cal D}(z)}{\Dfid(z)} \right] \; ,
\end{equation}
where
\begin{eqnarray}
{\cal D}(z) & \equiv & \int_0^z 
   \frac{dz^{\prime}}{{\cal H}[z^{\prime},\tau(z^{\prime})]} \nonumber \\
\Dfid(z) & \equiv & \int_0^z \frac{dz^{\prime}}
   {\sqrt{\Omato(1+z^{\prime})^3+\Olamo}} \; ,
\end{eqnarray}
and where the modified Hubble parameter ${\cal H}$ is specified as before
by Eq.~(\ref{newHub}).  To compare our predictions with the magnitude
residuals measured by Tonry et al. (2003) for 92 SNIa at $z>0.1$,
we write $\Delta m(z)=5(y\pm\delta y)-5\log [c(1+z)\Dfid(z)]$, where $z$,
$y$ and $\delta y$ are read from columns 7, 8 and 9 of Table~15 in that paper
[$y=\log (\dL\Ho)$].  To incorporate the new and invaluable survey of
23 SNIa at $z\sim1$ compiled by Riess et al. (2006), we note that
$y=\log (\dL\Ho)=\smallfrac{1}{5}(\muo-{\cal C})$ where $\muo$ is read from
column 3 of that paper and ${\cal C}$ is a constant whose value is fixed
by requiring that both samples give consistent values of $\dL$ for
SN1997ff at $z=1.755$, implying that ${\cal C}=15.825$.

\begin{acknowledgements}

J.O. thanks the members of the Gravity Probe Theory Group 
at Stanford University for useful discussions.  P.S.W. and B.M. acknowledge
support from N.S.E.R.C. and U.M.(C.) respectively.

\end{acknowledgements}

\end{document}